\documentclass[preprint,proceedings]{rmaa}

\SetYear{2002}
\SetConfTitle{Galaxy Evolution: Theory and Observations}

\title{Clues on the Evolution of Cluster Galaxies From The Analysis of
Their Orbital Anisotropies}

\author{
  A. Biviano\altaffilmark{1}
  P. Katgert\altaffilmark{2}
  T. Thomas\altaffilmark{2}
  A. Mazure\altaffilmark{3}}

\altaffiltext{1}{INAF -- Osservatorio Astronomico di Trieste,
  via G.B. Tiepolo, 11 -- 34131 Trieste, Italy.}
\altaffiltext{2}{Leiden Sterrewacht, P.O. Box 9513 NL-2300 RA
The Netherlands.}
\altaffiltext{3}{OAMP, LAM, Traverse du Siphon-Les trois Lucs  
                 13012 Marseille, France.}

\suppressfulladdresses

\listofauthors{A.~Biviano, P.~Katgert, T.~Thomas, \& A.~Mazure}
\indexauthor{Biviano, A.} 
\indexauthor{Katgert, P.}
\indexauthor{Thomas, T.} 
\indexauthor{Mazure, A.}

\addkeyword{Galaxies: clusters}
\addkeyword{Galaxies: kinematics and dynamics}
\addkeyword{Galaxies: evolution}

\begin{document}
\maketitle 

\boldabstract{We study the evolution of galaxies in clusters by the
  analysis of a sample of $\sim 3000$ galaxies, members of 59 clusters
  from the ESO Nearby Abell Cluster Survey (ENACS, Katgert et
  al. 1998), for which redshifts, $R$-band magnitudes, as well as
  morphologies are available (Thomas 2003; Biviano et al. 2002, B02
  hereafter, and references therein).}

In order to make the most efficient use of our data, we combine the 59
clusters into a single ensemble cluster as described in B02.  In the
ensemble cluster, after excluding galaxies in substructures, we find
that there are 4 cluster galaxy populations that must be distinguished
because they have different phase-space distributions: (i) the
brightest ellipticals, with $M_R \leq -22$ (using $H_0=100 \, \mbox{km
sec}^{-1} \mbox{Mpc}^{-1}$), (ii) the other ellipticals together with
the S0 galaxies (we refer to this class as $E+S0$ hereafter), (iii)
the early spirals (Sa--Sb), and (iv) the late spirals and irregulars
(Sc--Ir) together with the emission-line galaxies (ELG's).

About 2/3 of all cluster galaxies (outside substructures) belong to
the $E+S0$ class. The shape of the $E+S0$ velocity distribution
indicates that these galaxies move on nearly isotropic orbits, $\beta
\approx 0$ (see also van der Marel et al. 2000). We can therefore use
$E+S0$ as isotropic tracers of the cluster gravitational potential. We
solve the Abel and Jeans equations (see, e.g., Binney \& Tremaine
1987) using both a direct non-parametric approach, and the inverse
method described by van der Marel (1994). We find that a NFW (Navarro,
Frenk, \& White 1997) mass profile with
$r_s/r_{200}=0.25_{-0.10}^{+0.15}$ (68\% confidence limits) provides a
very good fit to our data.  We use this mass profile to estimate the
anisotropy profiles for the other three cluster galaxy populations,
using the method of Solanes \& Salvador-Sol\'e (1990).

We do not find any acceptable solution for the brightest ellipticals,
most likely because these galaxies do not fulfil the conditions for
the applicability of the {\em collisionless} Jeans equations. As a
matter of fact, the brightest ellipticals mostly sit at the bottom of
the cluster potential well, and move very slowly, if at all
(B02). They have probably been slowed down by dynamical
friction, and could have grown by mergers of other massive galaxies
(Brough et al. 2002).

We do find acceptable solutions for the early spirals. We cannot
exclude fully isotropic orbits for these galaxies, but the data taken
at face value indicate that {\em in the inner cluster region} they move on
radially-anisotropic orbits ($\beta \approx 0.6$).  Since there is
evidence that these galaxies evolve into S0's (Thomas \& Katgert
2003), it is possible that the early spirals that are still
visible near the cluster center are those that have managed to avoid
transformation, by the amplitude and direction of their velocities.

We also find acceptable solutions for the class of late
spirals + ELG's. The anisotropy is close to zero in the center, but
there are not many galaxies (if any) of this class there. For radii
$\geq 0.5 \, r_{200}$ the anisotropy grows almost linearly, reaching
$\beta \approx 0.6$ at a radius $\sim 1.5 \, r_{200}$, which is the limit
of our observational data. Such an anisotropy profile suggests that
the late spirals + ELG's are field galaxies infalling into the
cluster.  The lack of these galaxies in the central cluster region
suggests that they get transformed (into dwarf galaxies) or destroyed,
once they reach the high density central regions of the clusters.

\end{document}